\documentclass[preprint,nopacs,preprintnumbers,amsmath,amssymb]{revtex4}
\usepackage{graphicx,subfloat}
\begin{document}

\title{ Magnetoplasmons for the  $\alpha$-T$_3$ model with filled Landau levels}
\author{Antonios Balassis$^1$,  Dipendra Dahal$^2$,  \\   Godfrey Gumbs$^{2,3}$, Andrii Iurov$^{4}$,  Danhong Huang$^{5,6}$, and Oleksiy Roslyak$^1$}

\affiliation{$^1$ Department of Physics \& Engineering Physics, Fordham University,  441 East Fordham Road,  Bronx, NY  10458 USA\\
$^{2}$Department of Physics and Astronomy, Hunter College of the City University of New York, 695 Park Avenue, New York, NY 10065, USA\\
$^{3}$Donostia International Physics Center (DIPC), P de Manuel Lardizabal, 4, 20018 San Sebastian, Basque Country, Spain\\
$^{4}$ Medgar Evers College of the City University of New York, 1650 Bedford Avenue, Brooklyn, NY 11225 , USA\\
$^{5}$Center for High Technology Materials, University of New Mexico,
1313 Goddard SE, Albuquerque, New Mexico, 87106, USA \\
$^{6}$Air Force Research Laboratory, Space Vehicles Directorate,
Kirtland Air Force Base, New Mexico 87117, USA}

%
%

\begin{abstract}

\medskip

Using the $\alpha$-T$_3$ model, we carried out analytical and numerical calculations for the static and dynamic  polarization functions in the presence of a perpendicular magnetic field. These results were employed to determine the longitudinal dielectric function and the magnetoplasmon dispersion relation. The magnetic field splits the continuous valence, conduction and flat energy subband into discrete Landau levels which present significant effects on the polarization function and magnetoplasmon dispersion. We present results for a doped layer in the integer quantum Hall regime for fixed hopping parameter $\alpha$  and various magnetic fields as well as chosen magnetic field and different $\alpha$ in the random phase approximation.

\end{abstract}

\vskip 0.2in


\medskip
\par
\maketitle

\section{Introduction}
\label{sec1}

\begin{figure}
\begin{center}
\includegraphics[width=.55\textwidth]{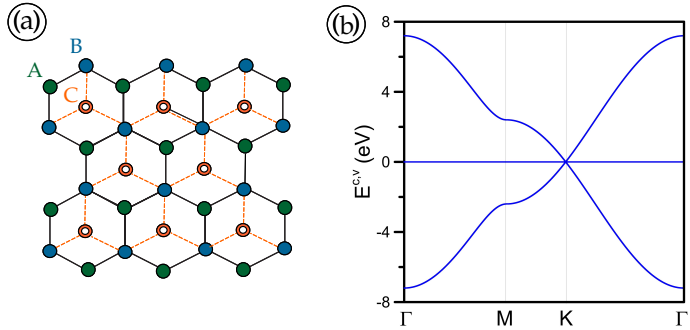}
\end{center}
\caption{(a) Schematic of the $\alpha$-T$_3$ lattice showing an atom C at the hub of a honeycomb structure having types A and B atoms on sublattices. Panel (b) shows the low-energy band structure  along high symmetry directions.   At low energy,  there exists a pair of linear  bands intersecting at the high symmetric K point because of the honeycomb symmetry along with a flat dispersionless subband.}
\label{FIG:1}
\end{figure}

In seminal work of Raoux, et al. \cite{Raoux}, it was demonstrated that Dirac cone structures  \cite{21,15,20,16,17,18,22,Bercioux,PRBAp2020} with the same energy band structure in the absence of magnetic field show substantial differences in their orbital magnetic susceptibilities. These range from diamagnetism in graphene \cite{gr01,gr02,gr03,ROSLYAK} to paramagnetism  in the T$_3$ or dice lattice \cite{Sutherland,thesis2,ABpaper,Vidal,Dey}.   The dice lattice, a sketch of which is  shown in Fig.\ \ref{FIG:1},  is defined by a Dirac-Weyl Hamiltonian similar to that for graphene, except that its pseudospin S = 1. The impact from Ref.\ \cite{Raoux} basically comes from its introduction  of a lattice parameter $\alpha$ which can be varied in a continuous way from the low-energy Dirac cone model to that for the dice lattice.  A unique property of this model is that the Berry phase can be varied  continuously from $0$ to $\pi$ by changing a parameter $\alpha$ which   represents the coupling strength between an additional atom at the center of the honeycomb graphene lattice and the A and B atoms of graphene, depicted in  Fig.\ \ref{FIG:1}(a).  Other properties of the $\alpha$-T$_3$ model which have been investigated include the magneto-optical conductivity and the Hofstadter butterfly \cite{AIinsert}, Floquet topological phase transition \cite{IND1}, the role of pseudospin polarization and transverse magnetic field on zitterbewegung  \cite{IND2}, its frequency-dependent magneto-optical conductivity \cite{hungary},  its magnetotransport properties  \cite{IND3} as well as the Hall quantization and optical conductivity \cite{thesis2}.   Also, the electron states of the gapped  $\alpha$-T$_3$ lattice in the presence of an electrostatic field of a charged impurity were reported recently \cite{gapped}.  We investigate the combined effect of varying $\alpha$ and a perpendicular magnetic field on the magnetoplasmon excitations of the $\alpha$-T$_3$ model. One possible realization of this model was given as cold atoms in an optical lattice\cite{Raoux}. Furthermore, there has been a proposal for its use as an optical lattice \cite{Dario1},  and it has been mentioned as having potential application to topology-induced phase transitions \cite{Dario2}.

\par

\medskip
\par
As is generally done for monolayer graphene, several authors have made a low-energy expansion of the band structure around the Dirac points  ${\bf K}= \left(\frac{4\pi}{3\sqrt{3}a_0}, 0\right)$ and ${\bf K}^\prime=\left(-\frac{4\pi}{3\sqrt{3}a_0}, 0\right)$ of the hexagonal Brillouin zone. In this notation, $a_0$  is the atom-atom lattice parameter. In this approximation, one can  investigate orbital susceptibility \cite{Raoux,Raoux2}, the feequency-dependent polarizability, impurity shielding, and plasmons \cite{Malcolm,hu, at1, malcolmMain}, Klein tunneling \cite{Klein}, and the  magnetotransport properties \cite{Biswas},   for the pseudospin-1 dice lattice.  In the low-energy regime, the energy subbands are given by $\epsilon({\bf k})=\pm\hbar v_F |{\bf k}|$, where $v_F$ is the Fermi velocity,  for the valence and conduction bands and a third {\em flat} band with zero energy, independent of the wave vector ${\bf k}$, as is represented in  Fig.\ \ref{FIG:1}(b). An interesting feature which the $\alpha$-T$_3$ model  exhibits is that a continuously variable Berry phase  does not change the energy band spectrum but some key physical properties  are strongly affected. However, this behavior is not maintained when there is a symmetry-breaking external  field.  Iurov, et al. \cite{Dressed}  investigated interacting Floquet states due to  off-resonant coupling of Dirac spin-1 electrons in the $\alpha$-T$_3$ model from external radiation having various polarizations. In particular, these authors demonstrated that when the parameter $\alpha$ is varied the electronic properties of the $\alpha$ -$T_3$ model (consisting of a flat band and two cones) could be  modified depending on the polarization of the external irradiation.  Furthermore, under elliptically-polarized light the low-energy band structure  depends on the valley index.

\medskip
\par 
It would be of interest to consider superfluidity and Bose-Einstein condensation  for dilute two-component dipolar excitons in $\alpha$-T$_3$. But, since this material is intrinsically gapless, we must find a way to open up a gap in order to separate the electrons in the conduction band from the holes in the valence band. This may be achieved by applying a perpendicular magnetic field  \cite{Berman,MLG}.  For this, one requires the electron-hole wave function, as it was done for graphene \cite{Fertig}, for which the electron is confined to one layer and the hole in the other layer with a dielectric material between them. This two-body problem could be treated  in terms of a two-dimensional  harmonic oscillator approximation and by employing either the Coulomb potential or taking appropriate screening effects into account using the Keldysh potential.    Consequently, a natural first step is to completely understand the eigenstate properties of electrons and holes in this spin-1 material in a magnetic field applied to a monolayer and their resulting collective magnetoplasmon properties so that these results could be  applied to a double layer with weakly interacting Bose gas of the dipolar excitons at low densities. There, one may assume that exciton-exciton dipole-dipole repulsion exists between excitons only for separations which exceed distances  between the exciton and the classical turning point. The distance between two excitons cannot be less than this distance.

\medskip
\par

 Another type of quantum material  consisting of a decorated honeycomb lattice is the Kagome lattice, discussed recently by Kane and Lubensky \cite{Kane}.    (For a review of artificial flat band systems, see Ref.\  \cite{Review}.)  This class of materials has been receiving considerable attention partly due to its  high degree of frustration and topological insulator properties \cite{Kag17,Sof1,Sof2,Sof3,Sof4,Sof5,PRA,PRL114,NJP14,PRL15}.    The lattice consists of  massless spring-like edges connecting massive hinge-like nodes.   Generally, one may consider the typical Kagome lattice  in a periodic reference configuration.  For hexagonal FeGe,,   the Fe atoms form a quasi-2D layered Kagome lattice, i.e.,  a 2D   network of trihexagonal tiling  on the atomic scale. and can be considered as  being representative of a Kagome metal.  Its  electronic structure  has a characteristic orbital selective Dirac fermions and extremely flat bands across the whole Brillouin zone.The two dispersive bands, in analogy with graphene, meet at two inequivalent Dirac points forming massless Dirac fermions.

\medskip
\par

The rest of this paper is organized as follows.  Section \ref{sec2} is devoted to a description of the low-energy Hamiltonian of the $\alpha$-T$_3$ model under a perpendicular magnetic field. There, we also present the energy eigenstates in the two valleys which are then employed in Sec.\ \ref{sec3} for calculating the form factors appearing in the polarizability. A thorough examination of the static polarization function at T=0 K is conducted. We have shown that as $\phi\to 0$ (or $\alpha\to 0$), the electronic excitation energy between a pair of specific Landau levels becomes proportional to $\sin\phi$ which tends to zero, and causes  the existence of a soft plasmon mode or instability in the   system through  the divergent behavior of the polarization function like $1/\sin\phi$. The  effect on those results due to finite frequency are also discussed in Sec.\ \ref{sec5}. We present our numerical results for magnetoplasmons corresponding to various coupling strengths and filling factors  in Sec.\ \ref{sec5}. Section\ \ref{sec7} is devoted to concluding remarks.

\section{Low-energy $\alpha$-T$_3$ Hamiltonian under a perpendicular magnetic field}
\label{sec2}

\subsection{Wave functions of the $\alpha$-T$_3$ lattice in the K valley}

 In the absence of an applied magnetic field, and with nearest-neighbor hopping in a single layer, the kinetic energy part of the  Hamiltonian  for the $\alpha$-T$_3$ model is obtained by including hopping contributions around the rim of the hexagon in a honeycomb lattice as well as from the hub atom to the rim.

\medskip
\par

For a chosen spin state, the degrees of freedom for the (A,B) sublattices and pseudospin ($\Uparrow,\Downarrow$) lead to four-component wave functions, which are written in the basis ($A\Uparrow,B\Uparrow,A\Downarrow,B\Downarrow$). In momentum space,  the  low-lying kinetic energy states near the Dirac point   for an electron in the absence of an applied magnetic field are given within the tight-binding approximation  by \cite{IND3,thesis2,thesis}

\begin{equation}
{\cal H}_s=\left(
\begin{array}{ccc}
{\cal H}_s^{(e)} & 0\\
0 & {\cal H}_s^{(h)}\\
\end{array}
\right)
\label{eqn:1}
\end{equation}
with $3\times 3$ submatrices for an electrons (e) and
holes (h)

\begin{equation}
{\cal H}_s^{(\ell)}=v_F \left(
\begin{array}{ccc}
0   &  \left(\tau \hat{p}_x^{(\ell)} +i \hat{p}_y^{(\ell)}   \right)\cos (\phi ) & 0 \\
 \left(\tau \hat{p}_x^{(\ell)} -i \hat{p}_y^{(\ell)}   \right)\cos (\phi )  &0   &  \left(\tau\hat{p}_x^{(\ell)} +i \hat{p}_y^{(\ell)}   \right)\sin (\phi ) \\
 0 &\left(\tau \hat{p}_x^{(\ell)} - i \hat{p}_y^{(\ell)}   \right) \sin (\phi )  &0  \\
\end{array}
\right) \ ,
\label{eqn:1eh}
\end{equation}
with $v_F$ denoting the Fermi velocity and where $\ell=e,h$ and $ \tau=\pm 1$ stands for the valley index at the $K(\tau=1)$ and $K^\prime(\tau=-1)$ points and $\phi$,  related to $\alpha$ as $\alpha=tan\phi$, represents the coupling strength.

\medskip
\par

We adopt the Hamiltonian describing the effects of a magnetic field $B_z\hat{z}$  perpendicular to the plane of the lattice  as derived in Ref. \cite{thesis}.  Since the spin-orbit coupling is weak, there is no buckling of the  structure such as occurs in phosphorene.  Consequently, magnetic field effects on $\alpha$-T$_3$ lattices are only manifested by perpendicular fields, whereby a Lorentz force on the charges is generated. We will  work in the Landau gauge, where the vector potential ${\bf A}$ is chosen so that ${\bf A}=-B_z y\hat{x}$  and $\nabla \times {\bf A}= B_z\hat{z}$ is the magnetic field. Using that Hamiltonian, one can calculate the wave functions and Landau levels for the lattice. In the Landau gauge for the vector potential ${\bf A}=-B_z y\hat{x}$ and using the usual Peierls substitution $\hbar{\bf k}\to {\bf p}\to {\bf p}+e{\bf A}$,  where ${\bf k}$  is the momentum  eigenvalue in the absence of magnetic field and ${\bf p}$ is the momentum operator,  we have

\begin{equation}
\hat{H}_K=-\hat{H}_{K'}^\ast=\gamma_B \begin{pmatrix}&&0 &&\cos\phi \ {\hat a} &&0\\
&&\cos\phi\ a^\dagger&& 0 &&\sin\phi\  {\hat a} \\ &&0 &&\sin\phi\ {\hat a^\dagger}&& 0\end{pmatrix}  \  ,
\label{Hamil}
\end{equation}
where $\gamma_B\equiv v_F\sqrt{2eB_z\hbar}=\hbar\omega_c$ with $\omega_c$ denoting the cyclotron frequency  and we introduced the destruction operator ${\hat a}=\frac{1}{\sqrt{2\hbar e B_z}}({\hat p_x}-e B_z {\hat y}-i {\hat p_y})$ and the creation operator ${\hat a^\dagger}=\frac{1}{\sqrt{2\hbar e B_z}}({\hat p_x}-e B_z {\hat y}+i {\hat p_y})$ analogous to the harmonic oscillator.   We note that when $\phi=0$, the Hamiltonian submatrix consisting of the first two rows and columns is exactly used in \cite{Roldan,Berman} for monolayer graphene. This important observation will come to have interesting consequences in this paper. The generalized wave function of this spin-1 Hamiltonian for $n\geq 2$ is given, in agreement with \cite{thesis},  by

\begin{equation}
\vert \Psi_{s}^{\tau}(\phi \, \vert \, {\bf k}) \rangle = \frac{1}{\sqrt{2}} \, 
\begin{pmatrix} \tau \left\{ \frac{ (2 n-\tau - 1) \, \cos^2 \phi }{ 2 n - \left[ 1 + \tau \cos(2 \phi)\right]  } \right\}^{1/2} \, \vert n - \tau - 1 \rangle \\
 s \, \vert n - 1 \rangle \\
 \tau \, \left\{
 \frac{ (2 n + \tau - 1) \, \sin^2 \phi }{ 2 n - 1 - \tau \cos(2 \phi) }
 \right\}^{1/2} \, \vert n + \tau -1 \rangle
\end{pmatrix}
   \frac{e^{i k_y y}}{L_y^{1/2}} \, ,
\label{VCeqn}
\end{equation}
where $s = \pm 1$ is the band index representing the  conduction $(s=1)$ and valence bands $(s=-1)$ with energy eigenvalue given by $\epsilon_s=s \gamma_B \sqrt{n-1+\sin^2\phi}$ with $n$ as Landau level index. If $n=1$, the first row in Eq. \ (\ref{VCeqn}) is zero  with energy eigenvalue $\epsilon_s=s \gamma_B\sin\phi$, as we discuss in detail below. And for the flat band $s=0$, the corresponding wavefunction is given compactly as
 
\begin{equation}
\vert \Psi_{s=0}^{\tau}(\phi \, \vert \, {\bf k}) \rangle = \,
\begin{pmatrix} \left\{ \frac{ (2 n+\tau - 1) \, \sin^2 \phi
 }{ 2 n - 1 - \tau \cos(2 \phi) } \right\}^{1/2} \, \vert n - \tau - 1 \rangle \\
 0 \, \vert n - 1 \rangle {\bf \equiv 0} \\
 - \, \left\{ \frac{ (2 n - 1 - \tau) \, \cos^2 \phi}{ 2 n - 1 - \tau \cos(2 \phi) }
 \right\}^{1/2} \, \vert n + \tau -1 \rangle
\end{pmatrix}
\frac{e^{i k_y y}}{L_y^{1/2}} \, ,
\label{FBeqn}
\end{equation}
where the graphene wave functions are

\begin{equation}
\Psi(x; n,k_y)\equiv
|n>=\frac{1}{\sqrt{2^n n! \pi l_H}}  \exp\left\{ -\frac{1}{2}\left(\frac{x-k_yl_H^2}{l_H}  \right)^2\right\}H_n \left(\frac{x-k_yl_H^2}{l_H}  \right) \ ,
\label{harmonic}
\end{equation}
expressed in terms of the Hermite polynomial $H_n(x)$, $k_y=2\pi m_y/L_y$ with $0\leq m_y\leq A/(2\pi l_H^2)$, $m_y$ is an integer, $L_y$ is a normalization length, $A$ is a normalization area and $l_H=\sqrt{\hbar /eB_z}$ is the magnetic length.

\medskip
\par

We note that for the K valley,    the eigenvalue of the lowest state $n=1$ has eigenvalue $\epsilon_s=s\gamma_B\sin\phi $  and eigenfunction

\begin{equation}
|\psi^K_{\pm,1}> =\frac{1}{\sqrt{2}}\begin{pmatrix} 0\\
\pm\ |0>\\
|1>\end{pmatrix} \ \frac{e^{ik_yy}}{L_y^{1/2}}  \  ,
\label{lowest1}
\end{equation}
which has different form in comparison to Eq. \ref{VCeqn}.  Also substituting $\phi=0$ in Eq. \  (\ \ref{VCeqn}) for the K valley, the two rows of Eq.\  (\ref{VCeqn}) give exactly the spin-\( \frac{1}{2}\) wave function for graphene. However, it is not the same as that for graphene due to the appearance of an additional element  \cite{Roldan}. This fundamental difference between the eigenstates of $\alpha-T_3$ model for $\phi=0$ and graphene is not restricted to the ground state but to all Landau levels.

\medskip
\par

  We also have from Eq.\ (\ref{FBeqn}), $n=0$  for the lowest flat band state as

\begin{equation}
|\psi^K_{\pm,1}> =\begin{pmatrix} 0\\
0\\
 \ |0>\end{pmatrix} \ \frac{e^{ik_yy}}{L_y^{1/2}}  \  .
\label{lowest2}
\end{equation}

\subsection{Wave functions of the $\alpha$-T$_3$ lattice in the K$^\prime$ valley}

The Landau levels for $K'$ valley are given by

\begin{equation}
\epsilon_s= s \gamma_B\sqrt{n-1+   \cos^2\phi}
\end{equation}
and the corresponding normalized eigenfunctions are obtained by replacing $\tau=-1$ in the Eq. \ref{VCeqn}.

\medskip
\par

Additionally, when $n=1$, the eigenfunction of the lowest state is
 
\begin{equation}
|\psi^{K'}_{\pm,1}> =\frac{1}{\sqrt{2}}\begin{pmatrix} -|1>\\
\pm\ |0>\\
0\end{pmatrix} \ \frac{e^{ik_yy}}{L_y^{1/2}}  \  ,
\label{lowest3}
\end{equation}
Now, setting $\phi=0$ in Eq. \ (\ref{VCeqn}),  the matrix element in the third row vanishes, the resulting eigenvector has three rows and does not coincide with the  pseudospin-$\frac{1}{2}$ wave function for graphene which has two rows only. Here the row with ket $|1>$ makes the difference with graphene.

\medskip
\par

By employing a  similar procedure to the one we followed above,  we obtain the normalized wave function for the flat band in the $K^\prime$ for $n\geq 2$ by replacing $\tau=-1$ in eq. \ref{FBeqn}.
For the flat band, the $n=0$, the wave function is

\begin{equation}
|\psi^{K'}_{0,0}> =\begin{pmatrix} |0>\\
0\\
0\end{pmatrix} \ \frac{e^{ik_yy}}{L_y^{1/2}}  \  .
\label{lowest4}
\end{equation}

\section{Polarizability for the $\alpha$-T$_3$ model}
\label{sec3}

A central quantity in our investigation is the frequency ($\omega$) and wave vector ($q$) dependent  longitudinal polarization function which  is generally given by
\
\begin{equation}
\Pi(q,\omega)=\sum_{s,s'}\sum_{n,n'}\frac{f(\epsilon_{s',n'})-f(\epsilon_{s,n})}{\hbar \omega+\epsilon_{s',n'}-\epsilon_{s,n}+i \delta}F^{\tau}_{sn,s'n'}(q) \ ,
\label{pol1}
\end{equation}
where $f(\epsilon_{s,n})$  is the Fermi-Dirac distribution function and $F^{\tau}_{sn,s'n'}(q) $ is a form factor which we  discuss below.  At zero temperature, we  have $f(\epsilon_{s,n})  = \theta(E_F-\epsilon_{s,n})$ in terms of the step function $\theta(x)$ and the Fermi energy $E_F$.    In this case,  the sums over the Landau levels in Eq. \ (\ref{pol1}) are cut off by a maximum number $N_F$, i. e.,  the filling factor. Additionally,   $N_F$ is related to the magnetic field  strength  through the relation, $N_F$=$\pi l_H^2 n_{2D}$  where $n_{2D}$ is the electron density and $l_H$    is the magnetic length. The Fermi wave vector $k_F$ is also given by $k_Fl_H$= $\sqrt{2N_F-1}$.

\medskip
\par

There are two distinct cases which we now address for the form factor corresponding to the allowed transitions from an occupied to an unoccupied state..

\vskip 0.2in

\noindent
{\em Case\ (i)}

The first corresponds to transitions between states within the conduction band (one below and the other above the Fermi level so that $ss^\prime=+1$) or from the valence to the conduction band so that $ss^\prime=-1$.  For these, the form factor is given compactly by

\begin{eqnarray}
&& F^{\tau}_{sn,s'n'}(q) \equiv |<\psi_{sn}^K|e^{i{\bf q}\cdot{\bf r}}|\psi_{s'n'}^{K}>|^2
\nonumber\\
&=&\frac{1}{4}\bigg|\begin{pmatrix} C_1(n)\cos\phi <n-2|&&\ s<n-1|&& C_2(n)\sin\phi  <n|\end{pmatrix}e^{i{\bf q}\cdot{\bf r}}\begin{pmatrix} C_1(n^\prime)\cos\phi  |n^\prime-2>\\  s^\prime |n^\prime-1>\\  C_2(n^\prime)\sin\phi  |n^\prime>\end{pmatrix}\bigg|^2
\nonumber\\
&=&\frac{1}{4}\left|C_1(n)C_1(n^\prime) R_{n-2,n'-2}\cos^2\phi +ss^\prime R_{n-1,n'-1}+C_2(n)C_2(n^\prime) R_{n,n'}\sin^2\phi \right|^2\ .
\label{27}
\end{eqnarray}
In our notation,
 
\begin{eqnarray}
C_1(n)  &=&\frac{\sqrt{n-1}}{\sqrt{n-1+\sin^2 \phi}}
\nonumber\\
C_2(n)&=&  \frac{\sqrt{n}}{\sqrt{n-1+\sin^2 \phi}} \ .
\end{eqnarray}

\begin{equation}
R_{n,n'}=\left<n|e^{i\bf q \cdot \bf r}|n'\right>=\sqrt\frac{2^{n_<} n_<!}{2^{n_>} n_>!}(-1)^{(n_> - n_<)/2}e^{iq_xk_y l_H^2}(q_xl_H)^{n_>-n_<}e^{-(q_xl_H/2)^2}L_{n<}^{n_>-n_<}\left(\frac{q_x^2l_H^2}{2}\right) \   ,
\end{equation}
where $L_n^m(x)$ is a Laguerre polynomial.   
It is worthy noting that when we set
$\phi=0$, in Eq.\ (\ref{27}),  the form factor takes the form as that obtained  in \cite{Roldan} for doped monolayer graphene  at T=0 in a perpendicular magnetic field.

\medskip
\par
 
\vskip 0.2in

\noindent
{\em Case\ (ii)}

\medskip
\par

It is now the turn for us to calculate the form factor for transitions from the $\epsilon=0$ (flat) band to the conduction band. 
The wave function for this degenerate level  is given in Eq.\ (\ref{FBeqn}) 
and that for the conduction band by Eq.\  (\ref{VCeqn}). Therefore,
the form factor for transitions from one of the discrete states with $\epsilon=0$ to the conduction band  is given by

\begin{equation}
F_{0,s'}^{n,n'}(q)=\frac{1}{2}\left|C_1(n')C_2(n)R_{n-2,n'-2}-C_1(n)C_2(n')R_{n,n'}\right|^2\sin^2\phi\cos^2\phi
\label{31}
\end{equation}
 Clearly,  this form factor vanishes when $\phi=0$, confirming that there is no contribution from the $\epsilon=0$ level states to the polarization for this case . This result is expected since when $\phi=0$, the $\alpha$-T$_3$ model yields the low-energy spin-$1/2$ graphene lattice.

In the next section, we present for the static polarization function at T=0 K  in Figs.\  \ref{fig1}, through  \ref{fig5}, Before doing so, we observe the following. 
For the form factors $F^{\tau}_{sn,s'n'}=\left|<\tau,s,n| e^{i\mathbf{q} \cdot \mathbf{r}}| \tau,s',n'>\right|^2$, where $\tau$ and $s$ are the valley and band indices, respectively, the following symmetry properties hold

\begin{eqnarray}
    F^{\tau}_{sn,s'n'}=F^{\tau}_{s'n',sn},\nonumber\\
    F^{\tau}_{sn,s'n'}=F^{\tau}_{-sn,s'n'}.\nonumber
\end{eqnarray}
Using these symmetry properties in the separation of the contributions to the polarization function in the way done for graphene \cite{Roldan}, we then obtain incorrectly that the flat band contribution to $\Pi(q,\omega)$ for any value of the parameter $\alpha=\tan\phi$ is zero.   Therefore, in doing our numerical calculations, we employed the form of the polarization function with the double Fermi-Dirac distribution function as given in Eq.\ (\ref{pol1}).

\medskip
\par

The first term in the summation for the  polarization function  in    Eq.\ (\ref{pol1})corresponds to $n=n'=0$ and, when $E_F=0$, it has the explicit form

\begin{equation}
\label{eqn:jump}
\Pi(q,\omega) \sim \frac{e^{-(ql_H)^2/2}}{\omega-\omega_c\sin \phi}\left (\frac{5}{16}(ql_H)^4-(ql_H)^2+1\right) \  ,
\end{equation}
which shows that in the long wavelength limit as the value of $\phi$ (and consequently $\alpha=\tan \phi$) is decreased, the contribution from this term is  increased and becomes infinite when $\phi=0$  (we set $\delta=0$). This equation explains the finite value (``jump'') as $q\to 0$ in Fig.\  \ref{fig2} and the large value of $\Pi(0.2,0)$ as $\alpha$ is decreased in Fig.\ \ref{fig4}(a).  We now elaborate on this unique behavior of $\alpha$-T$_3$  in magnetic field (finite $\omega_c$).

\medskip
\par

Our calculations thus show that under a perpendicular quantizing magnetic field, a series of Landau levels are generated in $\alpha$-$T_3$ lattices with level index $n=1,\,2,\,\cdots$,  but the $n=0$ Landau level coincides with the middle flat band at zero energy at the same time.  Due to broken time-reversal symmetry in the system, each original valley-degenerate graphene Landau level is split into two in $\alpha$-$T_3$ lattices with their energy separation depending on the values of the hopping parameter $\alpha$. In particular, the $K$-valley $n$-th Landau level approaches the energy of the $K^\prime$-valley $n$-th Landau level as $\alpha\to 1$. Meanwhile,  this same $n$-th Landau level merges with the $K^\prime$-valley $(n-1)$-th Landau level when $\alpha=0$.

\medskip
\par

Compared with graphene, we  have  also found two new $K$-valley $n=1$ Landau levels for $\alpha$-$T_3$ lattices, corresponding to the lower and upper Dirac cones,  respectively, and they become degenerate when $\alpha=0$  by merging with the zero-energy flat band.   Consequently, as $\phi\to 0$ (or $\alpha\to 0$), the electronic excitation energy between these two specific levels becomes proportional to $\sin\phi$ and goes to zero, which implies the appearance of a new optical-like plasmon mode that is absent in graphene. On the other hand, the existence of this new plasmon mode is connected to a divergent behavior (i.e., $\sim 1/\sin\phi$) in the polarization function as $\omega\to 0$.  Physically, this predicted divergence relates to the generation of a huge polarization filed by a very small in-plane electric field,  i.e., the occurrence of a switchable spontaneous static polarization field in the $\alpha$-$T_3$ lattices.  This further implies a possible magnetic-field facilitated phase transition in $\alpha$-$T_3$ lattices from a conventional dielectric material to a paraelectric one as $\alpha$ goes to zero.

\section{Dependence of polarizability:  on  magnetic field and coupling parameter}
\label{sec4}

\begin{figure}[!htb]
\centering
\includegraphics[width=0.75\textwidth]{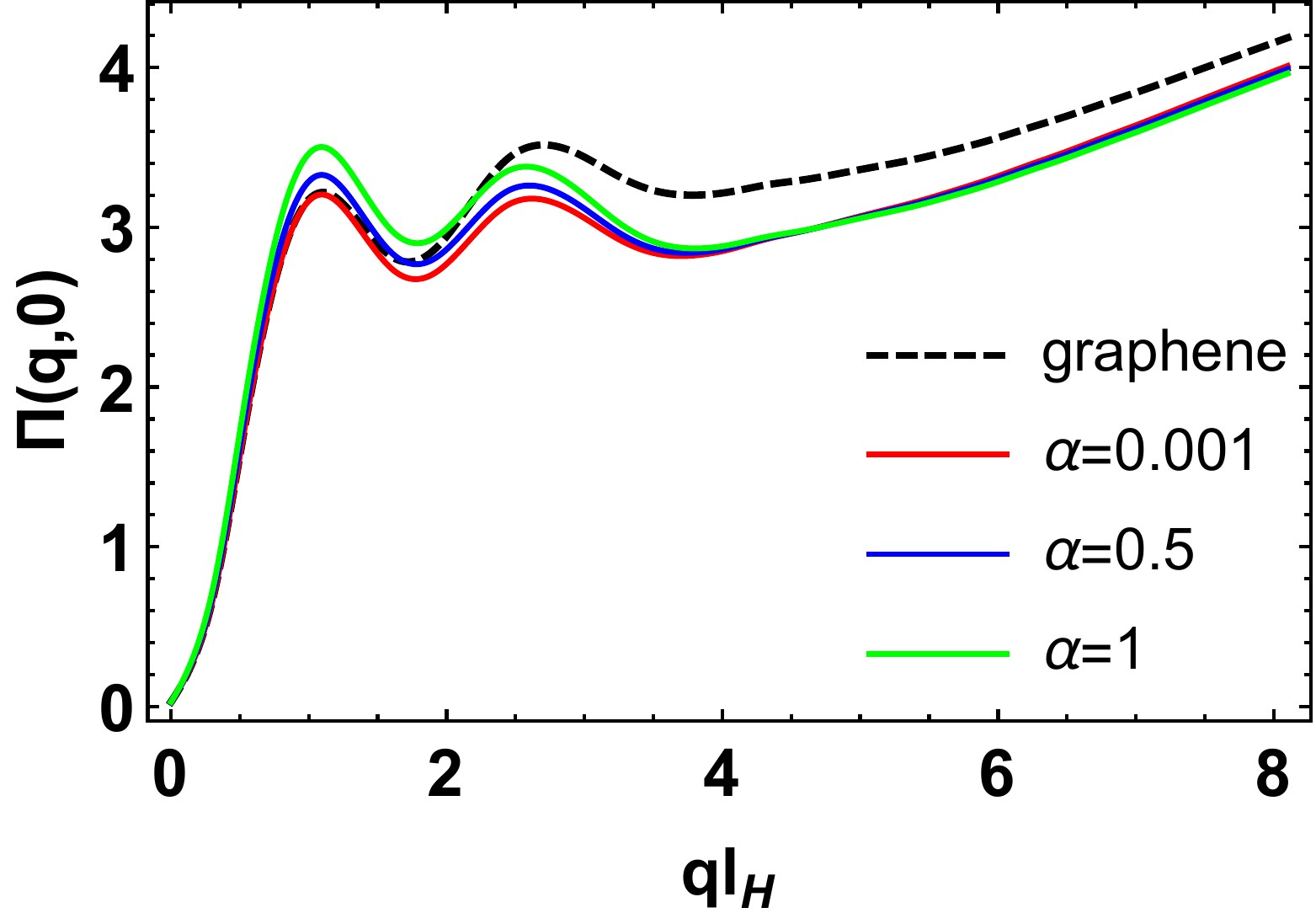}
\caption{Static polarization function $\Pi(q,0)$ vs wave vector $q$ for various values of the hopping parameter $\phi=\tan^{-1}\alpha$ for the case when the Fermi energy  $E_F$ corresponds to   the number of filled Landau levels $N_F=2$. The dashed line corresponds to graphene.}
\label{fig1}
\end{figure}

 In our plots for the magnetoplasmon dispersion relation, we used the cyclotron frequency  $\omega_c=\sqrt{2}\ v_{F}/l_{H}$ as the relative  scaling parameter and $r_s=e^{2}/\epsilon_{b} v_F=1$ for the interaction strength.  In  Fig.\  \ref{fig1}, we plot the static polarization for different values of the parameter $\alpha$ and fixed number of occupied states.

\medskip
\par
       
\begin{figure}[!htb]
\centering
\includegraphics[width=0.75\textwidth]{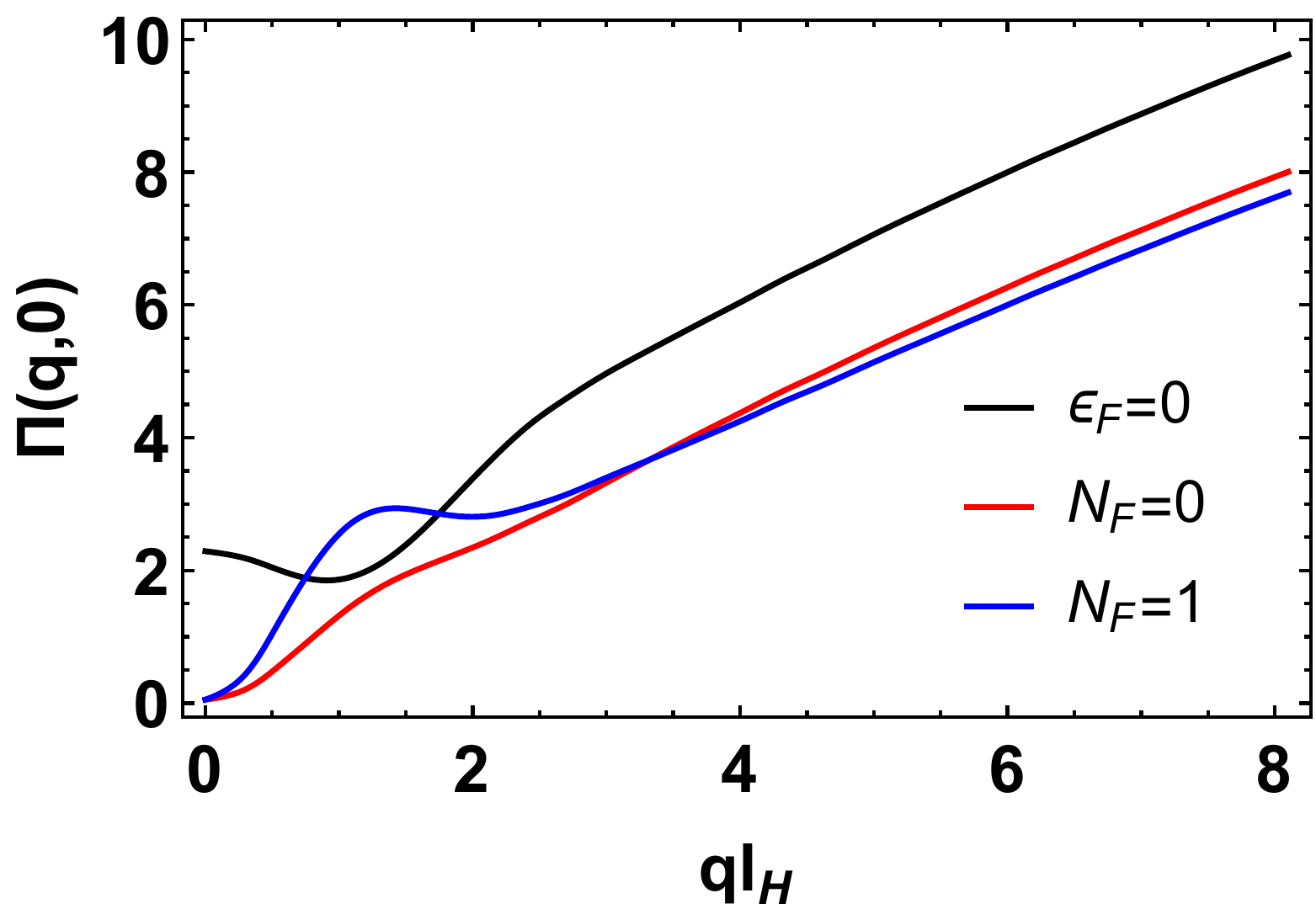}
\caption{(Color online)   Static polarization function $\Pi(q,0)$ vs wave vector $q$ for $\alpha=0.5$ and chosen Fermi energy $E_F$.  }
\label{fig2}
\end{figure}

 In Fig.\  \ref{fig2}, we present results for the static polarization function for chosen $\alpha=0.5$ and different numbers of occupied states.   We note that the ``jump'' described by Eq. \  (\ref{eqn:jump})    appears when the flat band is totally filled but the states above are empty.

\medskip
\par
       
\begin{figure}[!htb]
\centering
\includegraphics[width=0.75\textwidth]{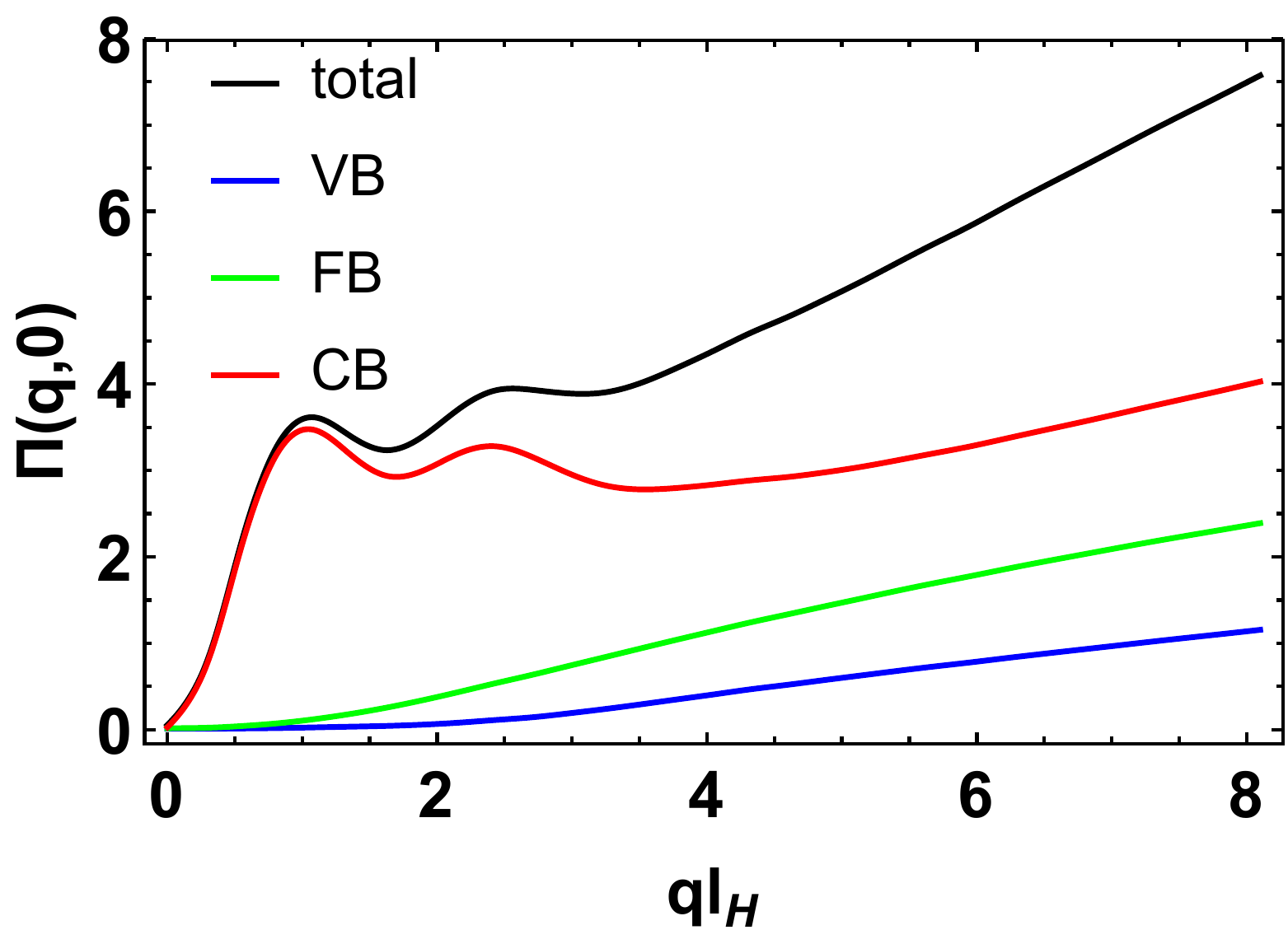}
\caption{(Color online)  Contributions to the total static polarization function $\Pi(q,0)$ (the black line) from the
valence band (VB), flat band (FB) and conduction band (CB) .
We chose  $N_F=0$ and $\alpha=1$, i.e., coupling parameter $\phi=\pi/4$..   }
\label{fig3}
\end{figure}

  In Fig.\  \ref{fig3}, we present the contributions to the total polarization (black line) due to transitions from different subbands.  These separate results are different for the valence band (blue line), flat band (green line), and the conduction band (red lines), for fixed Fermi level.

\medskip
\par

\begin{figure}
\centering
\includegraphics[width=0.75\textwidth]{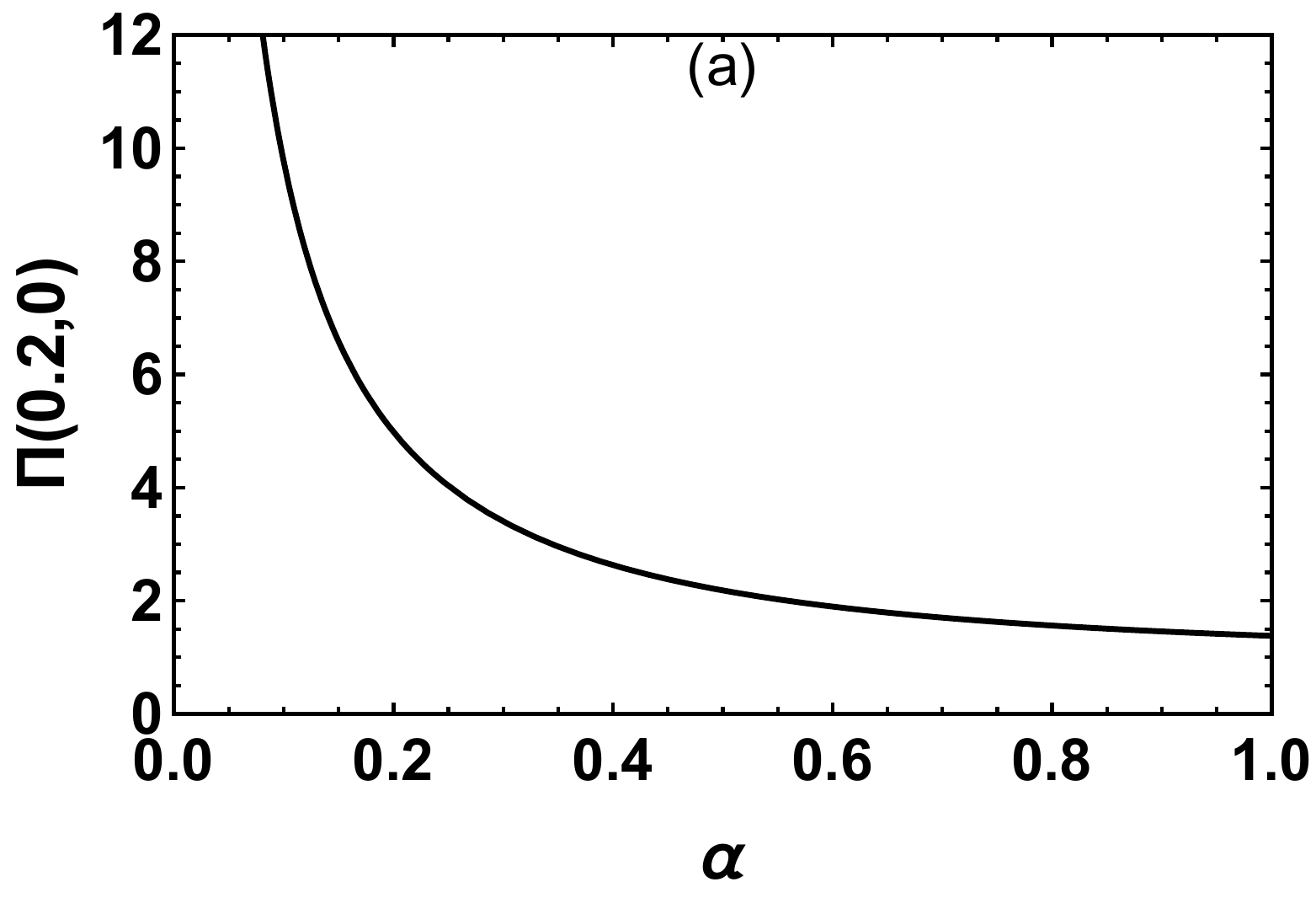}
\\
\includegraphics[width=0.75\textwidth]{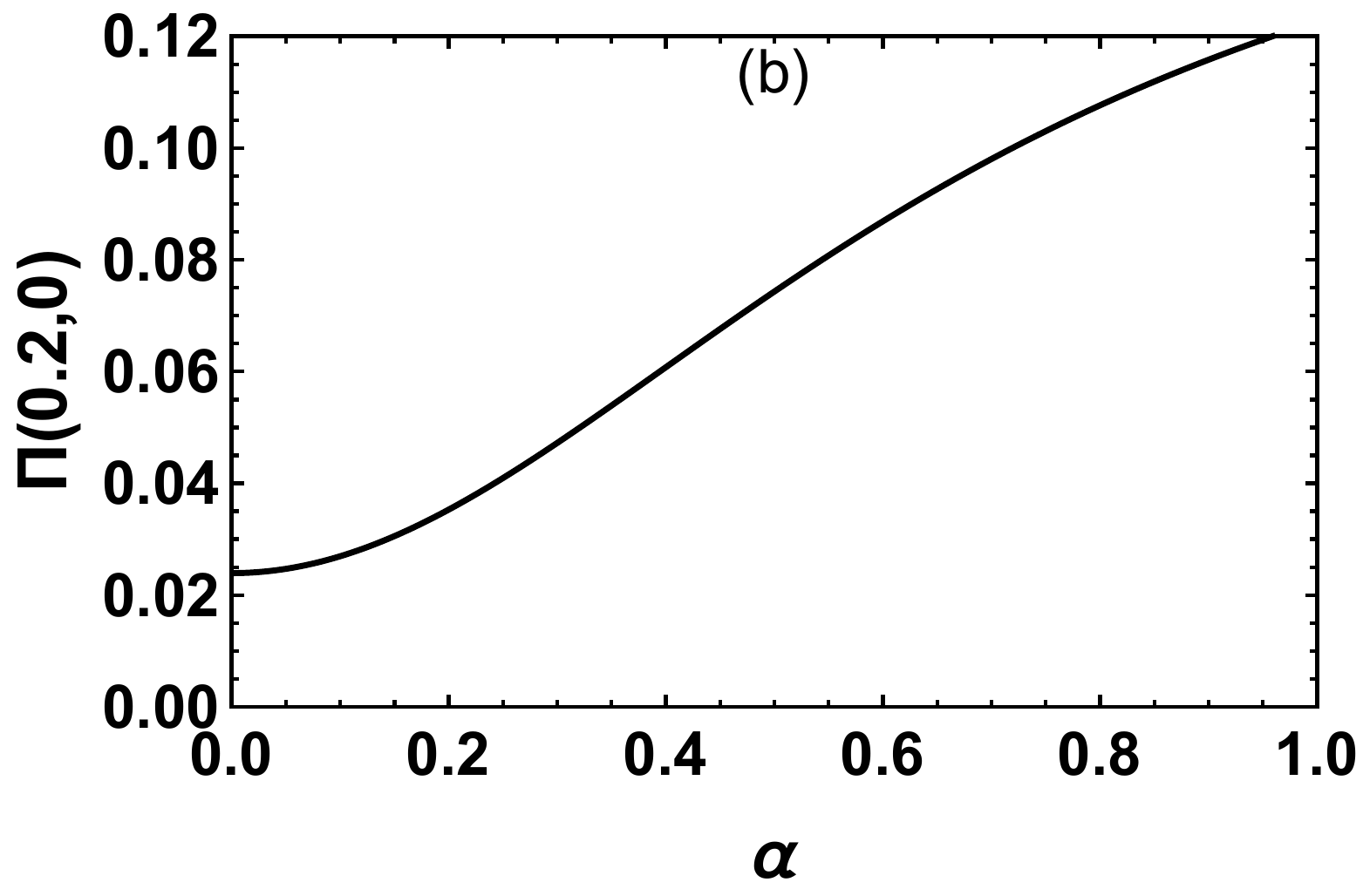}
\caption{Static polarization function vs coupling parameter  $\alpha=\tan \phi$ for chosen Fermi levels, (a) $E_F=0$ and (b)  $N_F=0$.}
\label{fig4}
\end{figure}

Figure\ \ref{fig4}  shows the way in which the  static polarization $\Pi(q,\omega)$ varies for fixed $ql_H=0.2$ as we vary the parameter $\alpha$. In Fig.\ \ref{fig4}(a),  the Fermi level is in the flat band whereas in Fig.\ \ref{fig4}(b), the Fermi level is chosen to be at the $n=1$ state of the conduction band. The large values of $\Pi(q,\omega)$ as $\alpha\to 0$ for the top Fig.\ \ref{fig4}(a), can  be explained from Eq.~(\ref{eqn:jump}).

\section{Magnetoplasmon Dispersion relation}
\label{sec5}

\begin{figure}[!htb]
\centering
\includegraphics[width=0.5\textwidth]{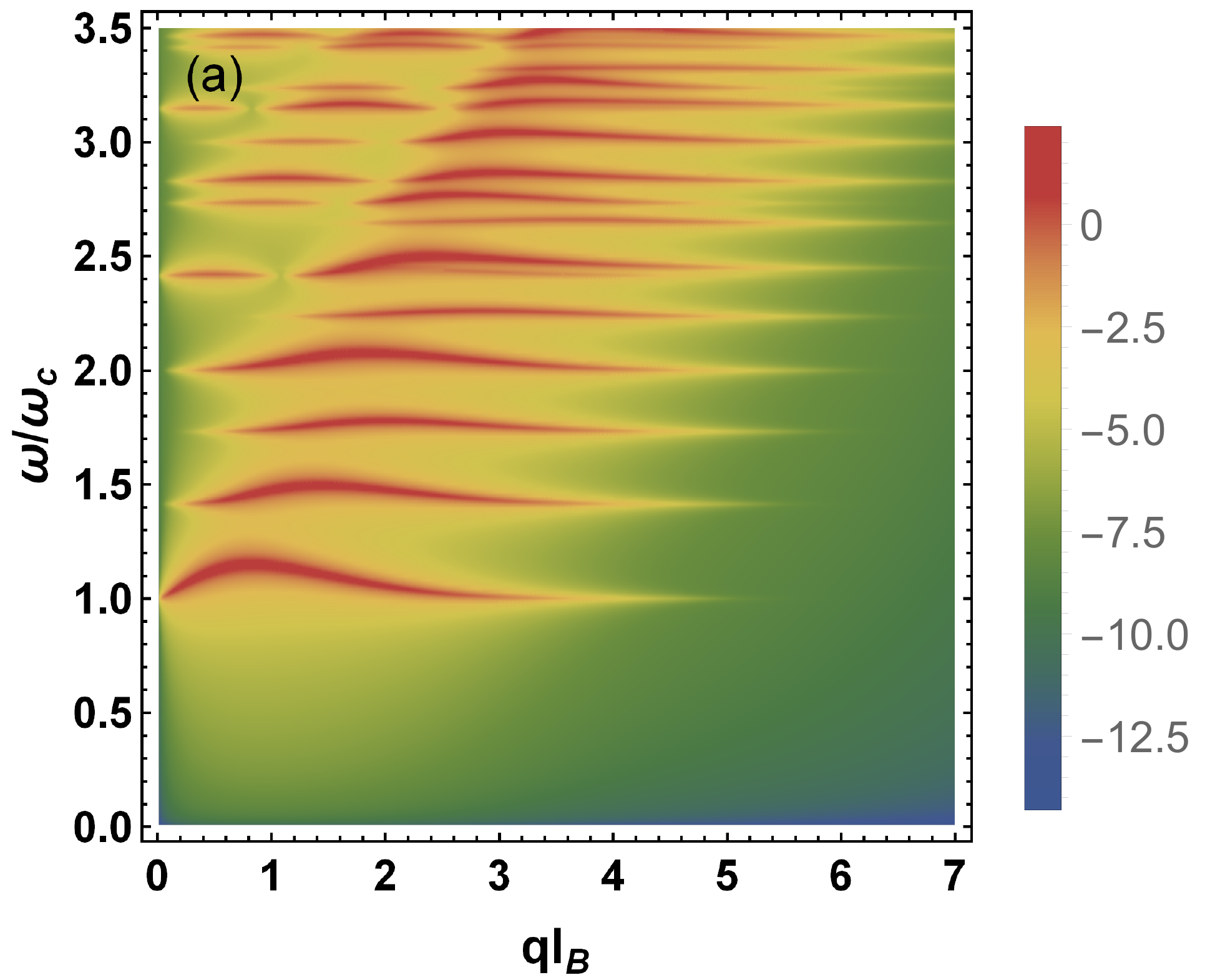}
\\
\includegraphics[width=0.5\textwidth]{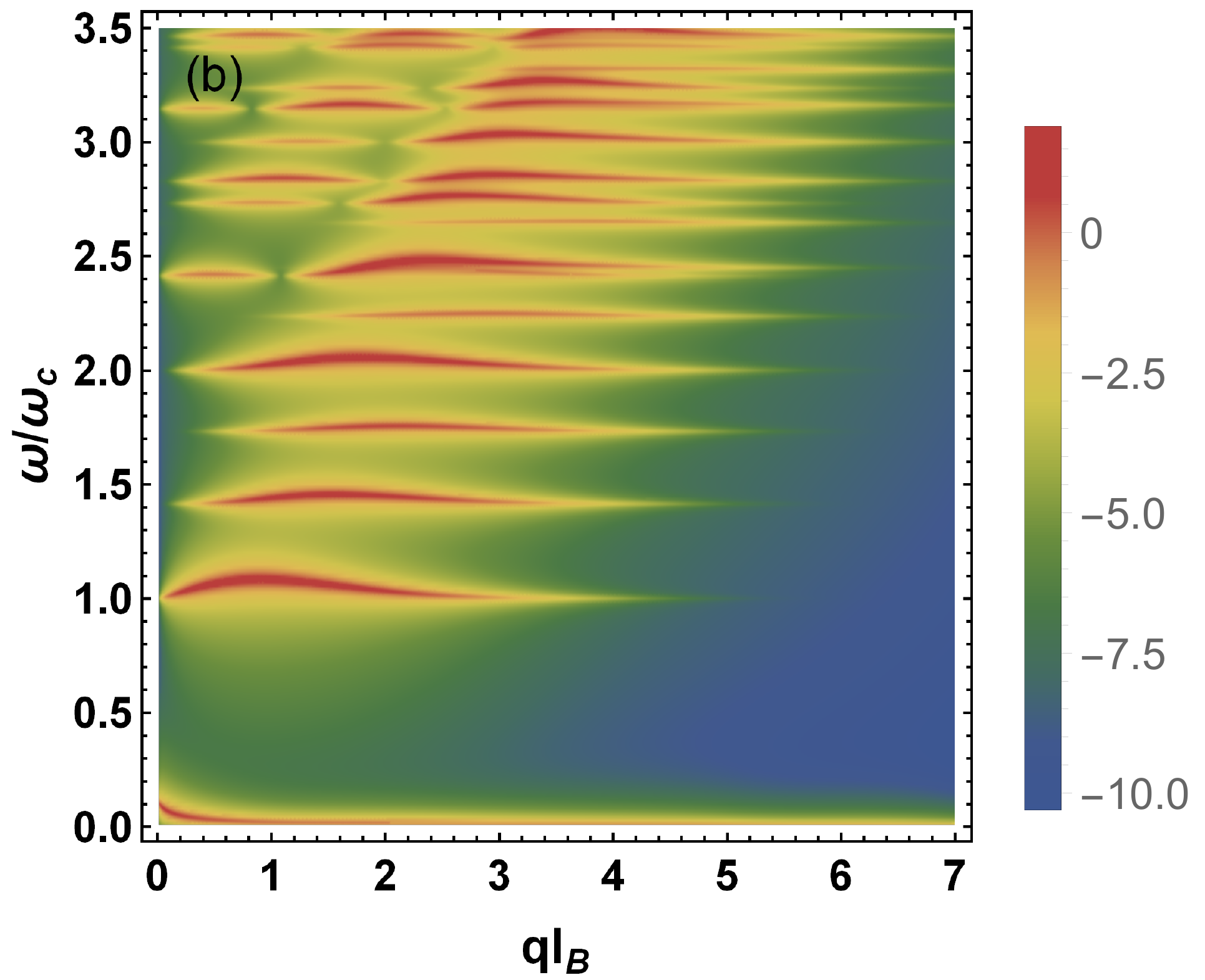}
\caption{(Color online)  Magnetoplasmon dispersion relation  for (a) graphene and (b) the $\alpha$-T$_{3}$ lattice when $\alpha=10^{-3}$ where the coupling parameter $\phi=\tan^{-1}\alpha$. The Fermi energy has been chosen to be $E_F=0$.}
\label{fig5}
\end{figure}

\begin{figure}
\makebox[\textwidth][c]{\includegraphics[width=0.7\textwidth]{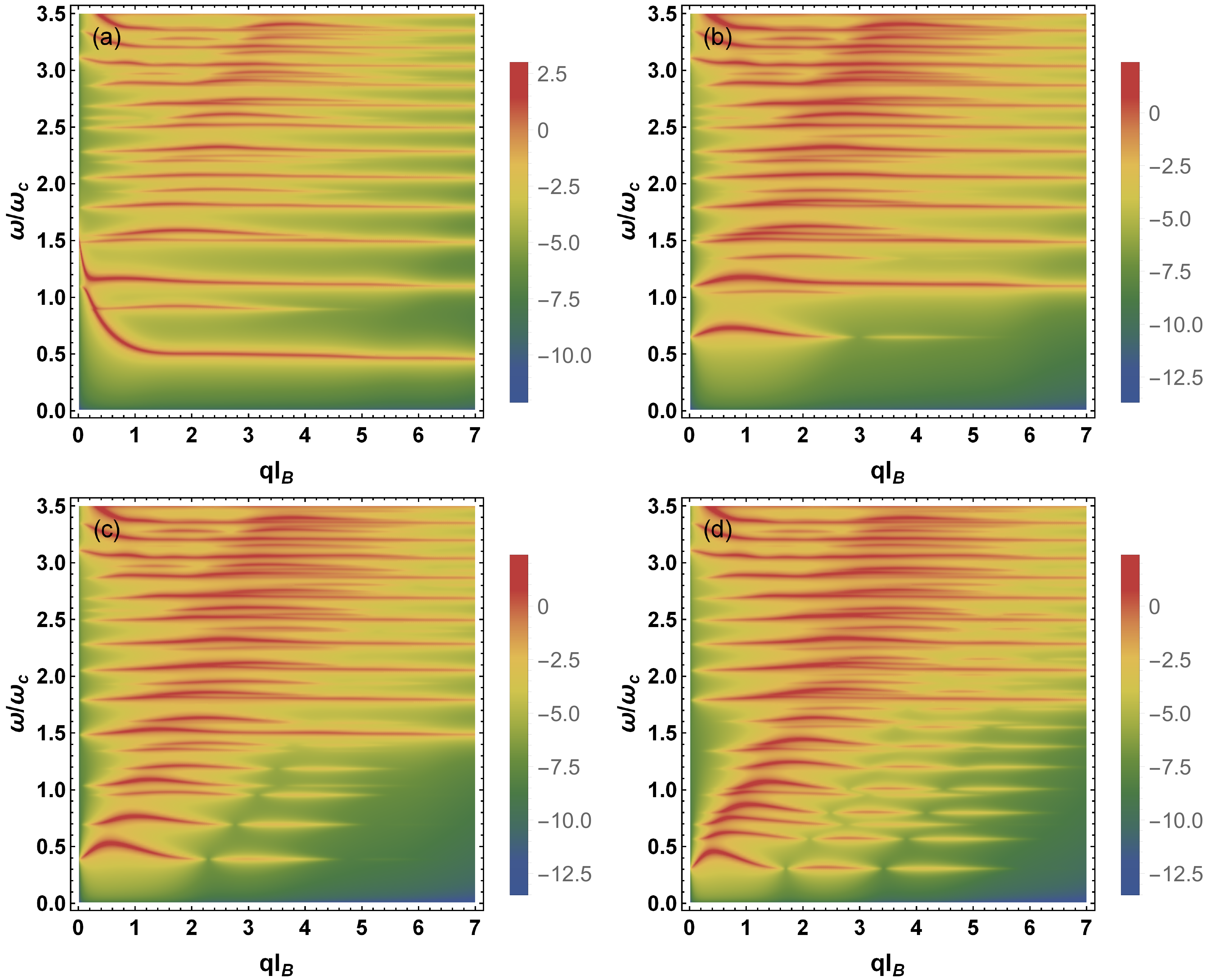}}
\caption{(Color online)  Magnetoplasmon dispersion relation  for $\alpha=0.5$, where the coupling parameter $\phi=\tan^{-1}\alpha$, and different Fermi energies. In (a) $E_F=0$ whereas in (b),  (c) and (d) the Fermi level was chosen  to lie in the conduction band with   (b)  $N_F=0$  ,(c)    $N_F=1$, and (d)   $n_F=2$..}
\label{fig6}
\end{figure}

Making use of the expression for polarization function in Eq.\ (\ref{pol1}), we have numerically calculated the plasmon mode dispersion relation for magnetoplasmons for the  $\alpha-{T_3}$ model in  the presence of a uniform perpendicular magentic field for various values of the coupling parameter  of $\alpha$. These correspond to the resonances of the polarizability for interacting electrons which, in the random-phase approximation (RPA), is

\begin{equation}
\Pi^{RPA}(q,\omega)=\frac{\Pi(q,\omega)}{1-v(q)\Pi(q,\omega)}
\equiv  \frac{\Pi(q,\omega)}{\epsilon(q,\omega)}\ ,
\end{equation}
where $v(q)$ is the Coulomb potential.  In Figs. \ \ref{fig5} and \ref{fig6}, we compare the dispersions of the $\alpha$-T$_3$ lattice  for various couplings represented by the choice for $\alpha$ and the Fermi energy $E_F$.

\medskip
\par

In Fig. \  \ref{fig5}, we show how the magnetoplasmon dispersion relation for graphene compare with those for the $\alpha-T_{3}$ lattice in the case when the Fermi level is chosen as $E_F=0$.  We mote the absence of the optical-like plasmon in graphene which appears at long wavelengths in the $\alpha$-T$_3$ in Fig. \  \ref{fig5}(b) which we prediced analytically in connection with behavior of the polarization function in Eq.\ (\ref{eqn:jump}).  In  Fig.\ \ref{fig6}, we plot the plasmon frequency versus wave vector for fixed $\alpha=0.5$ and different values of increasing Fermi energy. Comparing Fig.\ \ref{fig5}(b) with Fig.\ \ref{fig6}(a), we note that when we keep $E_F=0$ and we change $\alpha$ from zero to a finite value, the  form of the plasmon dispersion changes especially for low $\omega$ where the effect due to the Coulomb interaction is significant. From Fig.~\ref{fig6}, we see that for fixed value of the parameter $\alpha$, increasing the Fermi energy from zero to higher values changes the plasmon dispersion noticeably.

We can see several  bright branches originating at finite wave vector and the lower branches are generally more dispersive than  those at higher frequency.  Each magnetoplasmon branch is polarization shifted from a multiple of the cyclotron frequency $\omega_c$. In the lower frequency regime, these magnetoplasmons are of high intensity for longer wavelength but eventually fade away when the wave vector is increased due to Landau damping by single-particle excitations. Each branch bifurcates into less bright branches at larger wave vector and this division point shifts to larger wave vector in the higher frequency region.

\medskip
\par

 We also note that the frequency of the high-intensity portion of the low-energy magnetoplasmon branches increases monotonically and is then flattened for larger values of the wave vector where these lines are almost dispersionless. Another distinct feature seen in Figs. \ \ref{fig5} and \ref{fig6} is the minimum wave vector of the bright region for the magnetoplasmons. This critical wave vector is shifted to larger values for the higher branches in both Figs. \ \ref{fig5} and \ref{fig6}. This shift is relatively larger for the smaller values of $E_Fs$ can be seen in Fig. \ \ref{fig6}.    The brightness of the magnetoplasmons decreases drastically as the coupling parameter is decreased from $\alpha=1$.    Additionally, the overall intensity of the magnetoplasmons   is noticeably reduced for small $\alpha$.

\section{Concluding Remarks}
\label{sec7}
 
In summary, we have calculated the polarization function involving both analytical and numerical procedures and applied these results to a determination of the magnetoplasmon dispersion relation for the $\alpha$-T$_3$ lattice in the presence of perpendicular magnetic field.  Our numerical results are presented for the Landau levels at coupling strengths, expressed in terms of the parameter $\alpha$  between the hub atom and the atoms on the honeycomb lattice. In terms of Feynman diagrams,  the RPA utilizes the polarizability in the form of a particle-hole bubble so that mathematically, this is given by Eq.\ (\ref{pol1}).  Our numerical results for the zero temperature static polarizability at large wave vector show an interesting difference between the cases when the hopping parameter satisfies $0< \alpha\leq 1$ and the graphene case when $\alpha=0$. We emphasized that as $\alpha\to 0$, the plasmon mode softens due to a singularity in the polarization function which behaves like $\sim 1/\sin\phi$ as $\phi\to 0$.  From a physical point of view, we understand the difference in behavior between graphene and $\alpha$-T$_3$ as follows.

\medskip
\par

For $\alpha$-T$_3$ lattices, there are  two new $K$-valley $n=1$ Landau levels  corresponding to the lower and upper Dirac cones.  They become degenerate when $\alpha=0$.  Therefore,  in the limit when  $\phi\to 0$ (or $\alpha\to 0$), the electronic excitation energy between these two specific levels becomes proportional to $\sin\phi$ and vanishes  which  then yields  a unique optical-like plasmon mode  that is absent in graphene.  This predicted divergence generates  a large  polarization filed by a very weak in-plane electric field,  i.e., a spontaneous polarization field  is generated in the $\alpha$-$T_3$ lattices.  These results  further  suggest a possible magnetic-field facilitated phase transition in $\alpha$-$T_3$ lattices from a conventional dielectric material to a paraelectric one as $\alpha$ goes to zero.

\medskip
\par

The $\alpha$-T$_3$ lattice may be used in  electronic applications such as scattering control in  valleytronics since the wave function depends on the parameter $\alpha$. Its unique $\alpha$-dependent wave function may also be employed in coherent electro-optics, and its   edge-current in a nano-ribbon to control pseudospin-atom interaction.  Additionally, optical applications may arise from  band gap engineering to separate the flat band from the conduction (or valence) band. This could be achieved by controlled  light polarization for a topological transition.

\acknowledgments
 D.H. would like to acknowledge the support from the Air Force Office of Scientific Research (AFOSR). D.H is also supported by the DoD Lab-University Collaborative Initiative (LUCI) program.

\end{document}